\documentclass[a4paper,11pt]{article}
\usepackage{jinstpub}
\usepackage{lineno}
\usepackage{graphicx}
\usepackage{siunitx}

\title{\boldmath Calibration of the discriminator threshold of the Amplifier–Shaper–Discriminator Chip for the Upgrade of the ATLAS Muon Drift-Tube Chambers for the High-Luminosity LHC}

\author{O.~Kortner, N.~Meier, J.~Okfen}
\affiliation{Max-Planck-Institut für Physik,\\
Boltzmannstr. 8, 85748 Garching, Germany}

\emailAdd{Oliver.Kortner@mpp.mpg.de}

\abstract{
The read-out electronics of the ATLAS muon drift-tube chambers will be upgraded for operation at the High-Luminosity Large Hadron Collider. Part of the upgrade is the introduction of new amplifier–shaper–discriminator chips. This note presents the calibration of the discriminator threshold of the new chip in terms of primary electron signals using data from small-diameter muon drift-tube chambers operated in a monoenergetic, high-energy muon beam. The discriminator threshold intended for use in the ATLAS muon spectrometer corresponds to the signal of 20 primary electrons.
}

\keywords{Front-end electronics; Electronics for particle detectors; Muon detectors; Gaseous detectors; Drift chambers; Detector calibration methods; HL-LHC upgrade}

\begin{document}
\maketitle
\flushbottom

\section{Introduction}
\label{sec:intro}

The Large Hadron Collider (LHC) will be upgraded to the High-Luminosity LHC (HL-LHC) in the years 2026 to 2030, accompanied by upgrades of the LHC detectors. To cope with the tenfold increase in $pp$ collision rates from the LHC to the HL-LHC, the data acquisition of the ATLAS muon chambers will move to a streamed triggerless read-out, enabling the use of data from all detector technologies in the first-level trigger decision. Part of the upgrade of the read-out electronics of the ATLAS muon drift-tube chambers is a new Amplifier–Shaper–Discriminator (ASD) chip designed in 130~nm GlobalFoundries technology \cite{atlasMuonTDR, atlasTDAQTDR}.

In this publication, data from a small-diameter muon drift-tube chamber recorded in a 100~GeV muon beam at CERN are used to calibrate the discriminator threshold of the new ASD chip as multiples of single primary electron signals.

\section{The new ASD chip}
\label{sec:ASD}

\begin{figure}[hbt]
\begin{center}
    \includegraphics[width=0.75\linewidth]{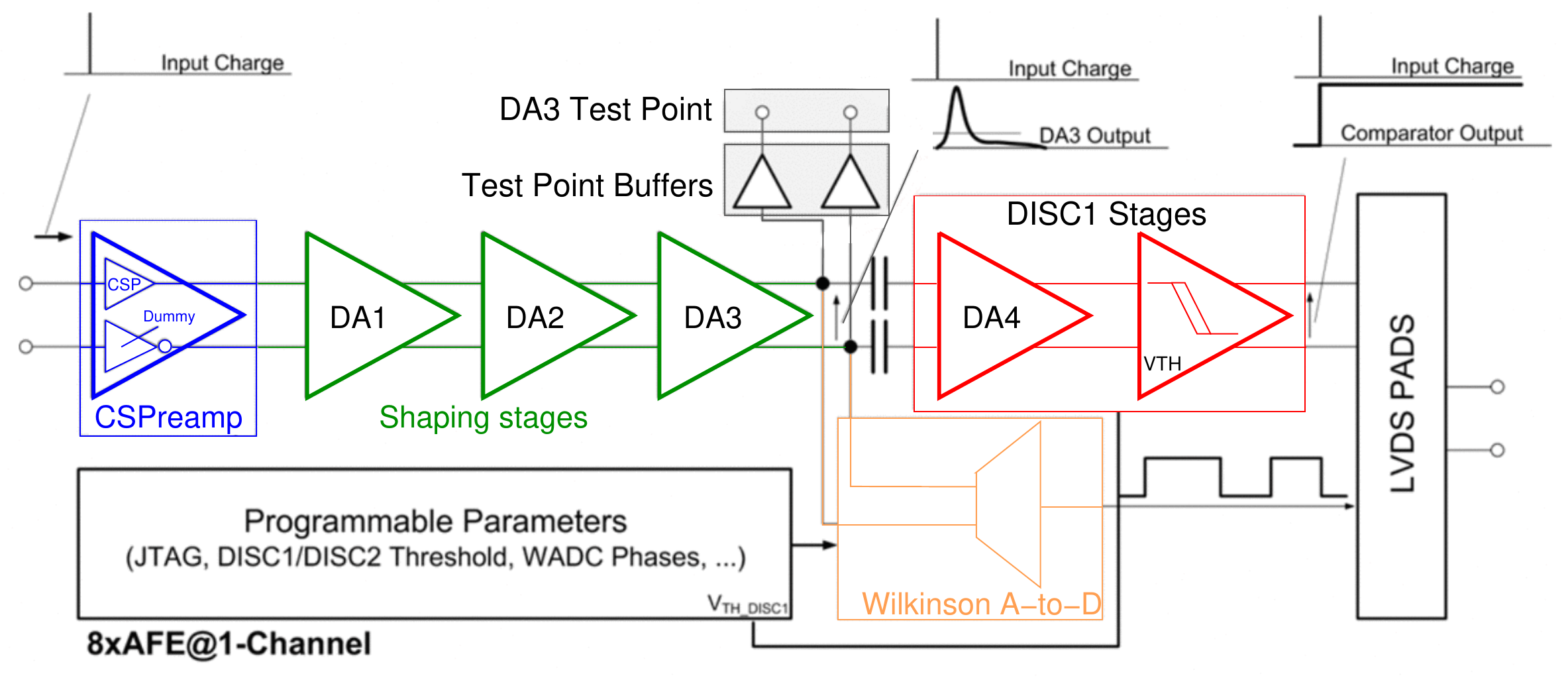}
    \caption{\label{fig03}Block diagram of the new ASD chip. \cite{ASDIEEE}}
\end{center}
\end{figure}

The ASD chip used in ATLAS at the LHC was designed in Agilent 500~nm technology, which has become obsolete. For the upgrade of the ATLAS muon spectrometer, a new ASD chip was designed in GlobalFoundries 130~nm CMOS technology \cite{ASDIEEE}. The design of the new chip follows that of the 500~nm chip but incorporates a correction of a specific design error in the output logic. A block diagram of the new ASD chip is shown in Figure~\ref{fig03}. The chip features a differential charge-sensitive preamplifier, bipolar shaping with ion-tail cancellation, and a Wilkinson ASD for time-walk corrections to the discriminated signals.

The discriminator is implemented as a Schmitt trigger, so the actual threshold is determined by two parameters: the discriminator threshold and the hysteresis. Both are programmed via JTAG, using an 8-bit word with the least significant bit corresponding to 3~mV for the discriminator threshold and a 4-bit word with the least significant bit corresponding to 2.4~mV. Values of the discriminator code between 0 and 126 correspond to negative voltages; values above this range correspond to positive voltages.

A relationship between the threshold and hysteresis settings and the actual threshold expressed as multiples of the primary electron signal is derived from muon beam data in the following sections.

\section{Experimental setup}
\label{sec:setup}

\begin{figure}[hbt]
\begin{center}
    \includegraphics[width=0.7\linewidth]{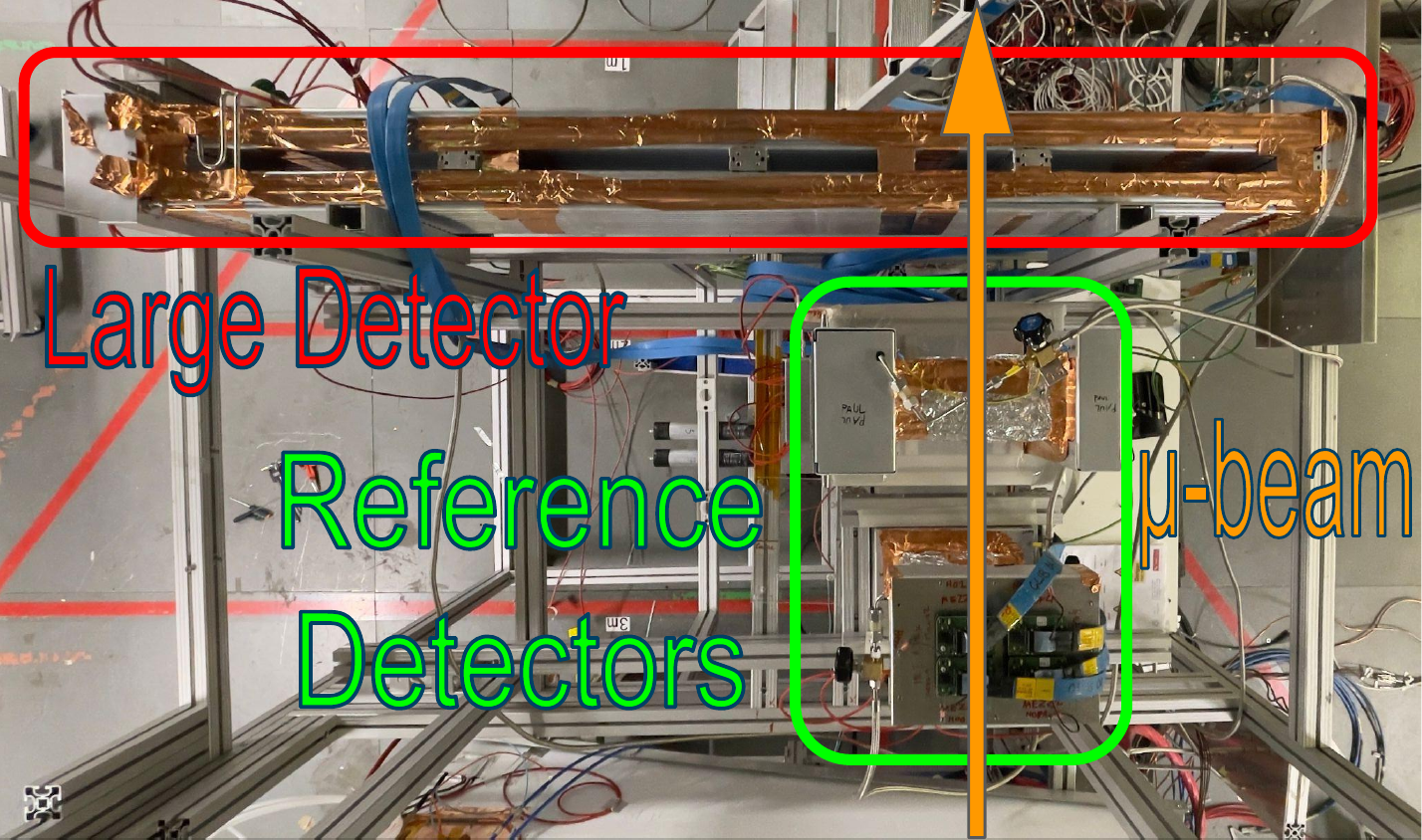}
    \caption{\label{fig:testbeam_setup}
    Photograph of the sMDT chamber setup at the GIF++ facility. The setup
    consisted of two small reference chambers and a large sMDT chamber used
    for the ASD parameter scans.}
\end{center}
\end{figure}

The setup consisted of two small-diameter muon drift-tube (sMDT) chambers used as reference trackers and one large sMDT chamber in which the ASD parameters were scanned. All three chambers were equipped with the new ASD front-end electronics. A photograph of the experimental setup is shown in Fig.~\ref{fig:testbeam_setup}.

The chambers were placed sequentially along the beam direction. The faces of two adjacent chambers were separated by approximately 25~cm. The first reference chamber was oriented to measure in the $x$-coordinate direction. The second reference chamber was rotated by 90$^\circ$ relative to the first and measured in the $y$-coordinate direction. The large test chamber was placed downstream of the reference chambers and oriented parallel to the second reference chamber, such that it also measured in the $y$-coordinate. The orthogonal orientation of the reference chambers provided track position information in both coordinates.

Each reference chamber consisted of eight tube layers with 12 drift tubes per layer, for a total of 96 tubes per chamber. The reference chambers used sMDT tubes with a length of 0.25~m, an outer diameter of 15~mm, and an aluminium wall thickness of 0.4~mm.

The large chamber used for the ASD measurements consisted of two multilayers separated by approximately 4~cm. Each multilayer comprised four tube layers with 12 tubes per layer. The chamber contained 96 drift tubes in total. The tubes had a length of 1.6~m, the same outer diameter of 15~mm, an aluminium wall thickness of 0.4~mm, and the same wire type as that used in the reference chambers.

All chambers were operated with an Ar/CO$_2$(93/7) gas mixture at an absolute pressure of 3~bar. The two reference chambers were operated at a fixed high voltage of 2730~V, corresponding to the nominal ATLAS gas gain of approximately $G_{\mathrm{nominal}} = 2 \times 10^4$. Their high-voltage settings, discriminator threshold, and hysteresis were kept constant throughout the measurements to provide stable reference-tracking information.

\section{Calibration of the discriminator threshold}
\label{sec:calibration}

To determine the actual threshold in terms of primary electron signals, the dependence of the efficiency on the high voltage applied to the drift tube was measured for different ASD threshold and hysteresis settings. The relationship between the gas gain $G$ and the operating voltage for the Ar/CO$_2$(93/7) gas mixture can be described by
\begin{eqnarray}
    \ln G = \frac{r_{wire}E(r_{wire})\ln2}{\Delta V}
            \ln\frac{E(r_{wire})}{E_{min}(\rho_0)
            \frac{\rho_{gas}}{\rho_0}}
    \label{sec:setup:eqn1}
\end{eqnarray}
with the radius of the anode wire $r_{wire}=25~\mu$m, the electric field on the surface of the anode wire $E(r_{wire})$, $\rho_{\mathrm{gas}/0}$ denoting the gas density at the operating conditions of the chamber and at normal conditions, respectively, and $E_{min}=24$~kV/cm, $\Delta V=34$~V \cite{Aleksa1999}. If $n_{prim}$ is the number of primary electrons produced by a muon in the gas and $G_{50\%}$ is the gas gain at which the muon efficiency equals 50\%, then the discriminator threshold at the nominal gas gain $G_{nominal}$ corresponds to
\begin{eqnarray}
    n_{prim}\cdot\frac{G_{50\%}}{G_{nominal}}.
    \label{sec::setup:eqn2}
\end{eqnarray}

This formula assumes that the signal electrons arriving at the anode wire do so within a time interval of the order of the peaking time of the amplifier. This condition is fulfilled by considering only muon tracks with at least 5~mm distance from the anode wire. The value of $n_{prim}$ was obtained from simulations using Garfield$^{++}$ \cite{GarfieldPP}. The most probable value was found to be $165.6\pm0.6$.

\begin{figure}[hbt]
\begin{center}
    \includegraphics[width=0.45\linewidth]{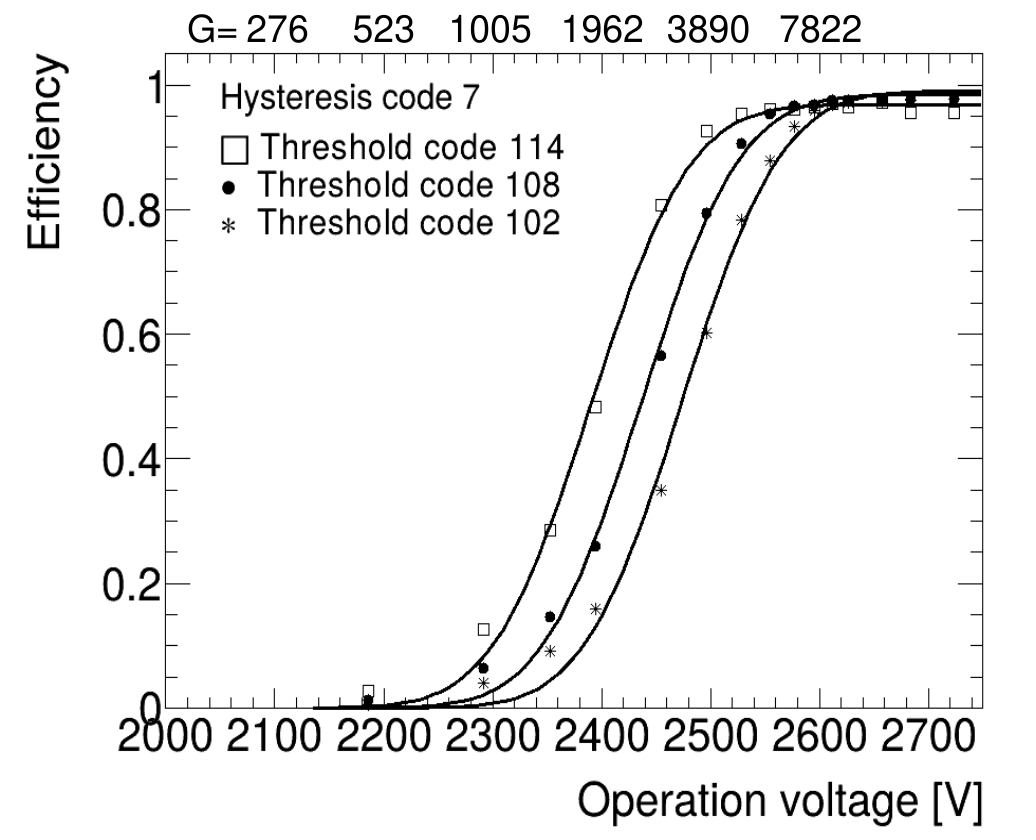}
    \includegraphics[width=0.45\linewidth]{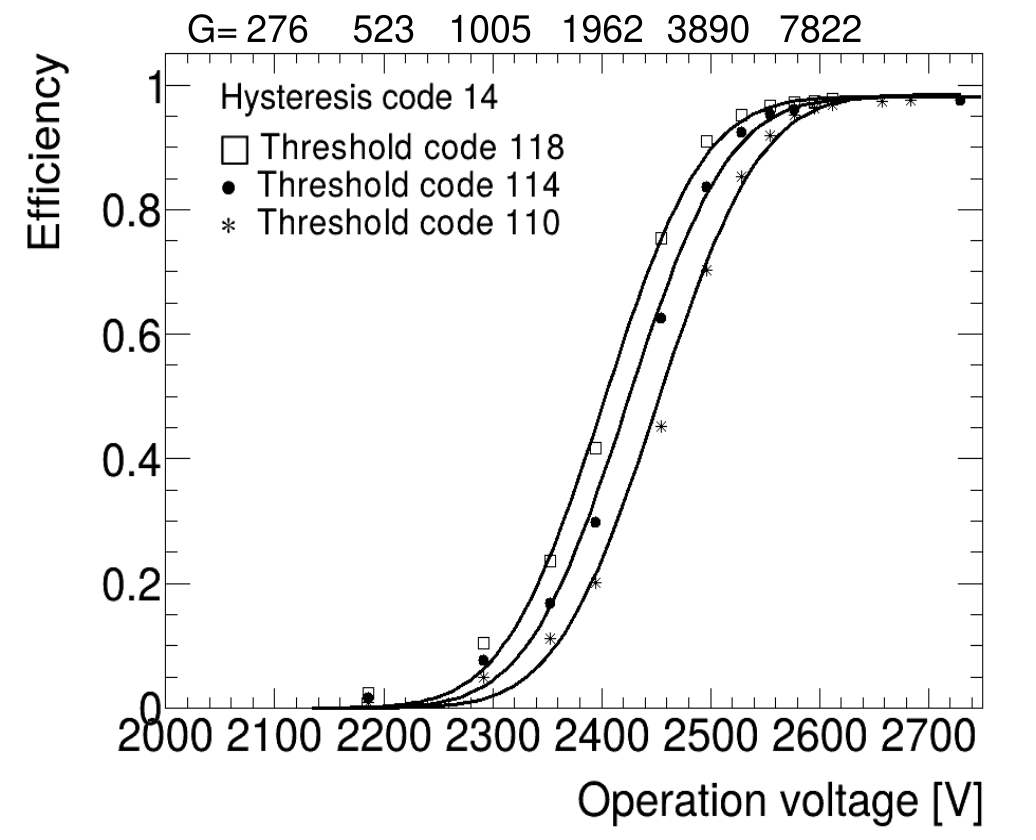}
    \caption{\label{sec:calibration:fig01}Measured dependencies of muon detection efficiencies of an sMDT on the operating voltage for different threshold and hysteresis settings. The gas gain obtained at the different voltages is displayed at the top. The lines represent fits of an error function to the measured efficiency points.}
\end{center}
\end{figure}

Figure~\ref{sec:calibration:fig01} shows the tube efficiencies measured at different operating voltages and different hysteresis and threshold settings in the large chamber. The settings of the reference chambers were not altered. The efficiency is defined as the ratio of the number of tubes traversed by the muon that show a hit to the total number of tubes traversed. Tubes are considered to be traversed if the distance of the reference muon track from the tube’s anode wire is less than 7~mm. To determine the point of 50\% of the maximum efficiency, error functions are fitted to the measured efficiency points.

\begin{figure}[hbt]
\begin{center}
    \includegraphics[width=0.5\linewidth]{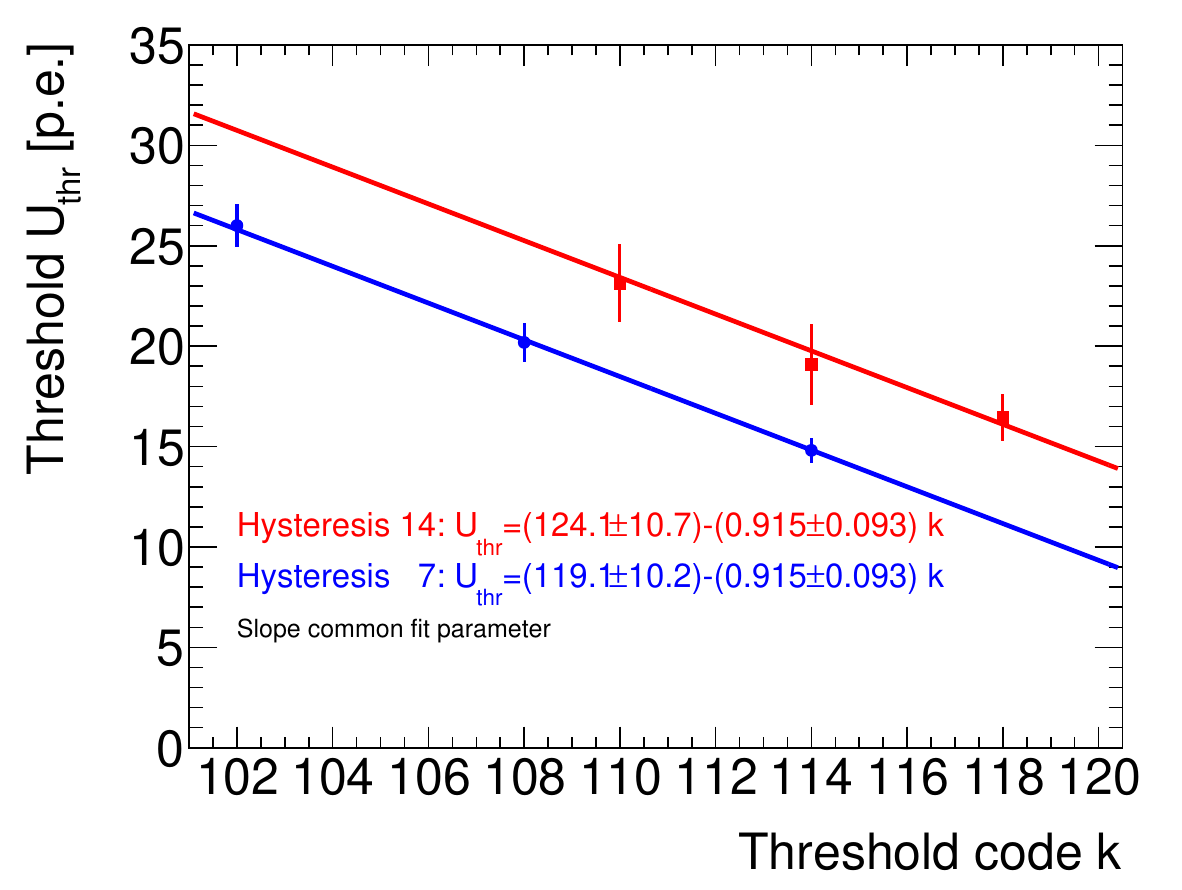}
    \caption{\label{sec:calibration:fig02}Dependence of the discriminator threshold in multiples of primary electron signals on the hysteresis and discriminator settings. The error bars include statistical and systematic uncertainties. The two fit curves are obtained using a common slope parameter.}
\end{center}
\end{figure}

The obtained 50\% efficiency points are converted into threshold values in terms of multiples of primary electron signals using Equation~(\ref{sec::setup:eqn2}). To estimate systematic uncertainties, the efficiencies are re-evaluated by considering muon tracks with distances between 5 and 7~mm from the anode wire. A systematic uncertainty between 0.5 and 1 primary electron is found. The results are shown in Figure~\ref{sec:calibration:fig02}. The fitted straight lines are determined with a common slope parameter. A change of the threshold code by 1 corresponds to a change of the actual discriminator threshold by 0.92 primary electrons. A change of the hysteresis by 1 corresponds to a change of the actual discriminator threshold by 0.71 primary electrons. The ratio of 0.92 to 0.71 primary electrons agrees with the ratio of the least significant bits of the threshold and hysteresis codes within uncertainties.

\section{Conclusions}
\label{sec:conclusions}

In this note, the calibration of the discriminator threshold of the amplifier–shaper–discriminator for the ATLAS muon chambers at the HL-LHC is presented. The ATLAS collaboration plans to operate the front-end electronics with a discriminator threshold corresponding to the signal of 20 primary electrons. According to the results presented here, this can be achieved with a threshold code of 108 and a hysteresis code of 7. If the hysteresis is changed by a value $\delta h$, the threshold code must be adjusted by $1.28\,\delta h$.

\end{document}